\documentclass[final,5p,times,twocolumn]{elsarticle}

\usepackage{amssymb}

 \biboptions{sort&compress}

\usepackage{float}
\usepackage{subfig}
\usepackage[export]{adjustbox}
\usepackage{placeins}
\usepackage{booktabs}
\newcommand{\head}[1]{\textnormal{\textbf{#1}}}
\usepackage{amsmath,bm}
\usepackage{pdflscape}
\usepackage{url}
\usepackage{textgreek}
\usepackage[final]{microtype}
\usepackage{multirow}

\usepackage[usenames]{color}

\usepackage{adjustbox}

\newcommand{\beginsupplement}{%
        \setcounter{table}{0}
        \renewcommand{\thetable}{S\arabic{table}}%
        \setcounter{figure}{0}
        \renewcommand{\thefigure}{S\arabic{figure}}%
        \setcounter{section}{0}
        \renewcommand{\thesection}{S\arabic{section}}
     }

\journal{Powder Technology}

\makeatletter
\def\ps@pprintTitle{%
 \let\@oddhead\@empty
 \let\@evenhead\@empty
 \def\@oddfoot{}%
 \let\@evenfoot\@oddfoot}
\makeatother

\begin{document}

\begin{frontmatter}

\title{Quantitative analysis of thin metal powder layers via transmission X-ray imaging and discrete element simulation: Roller-based spreading approaches}

\author[MIT]{Ryan~W.~Penny\corref{cor1}}
\ead{rpenny@mit.edu}
\author[MIT]{Daniel~Oropeza}
\author[MIT]{Reimar~Weissbach}
\author[TUM]{Patrick~M.~Praegla}
\author[TUM]{Christoph~Meier}
\author[TUM]{Wolfgang~A.~Wall}
\author[MIT]{A.~John~Hart\corref{cor1}}
\ead{ajhart@mit.edu}

\address[MIT]{Department of Mechanical Engineering, Massachusetts Institute of Technology, 77 Massachusetts Avenue, Cambridge, 02139, MA, USA}
\address[TUM]{Institute for Computational Mechanics, Technical University of Munich, Boltzmannstra{\ss}e 15, Garching b. M{\"u}nchen, Germany}

\cortext[cor1]{Corresponding author}

\begin{abstract}
A variety of tools can be used for spreading metal, ceramic, and polymer feedstocks in powder bed additive manufacturing methods. Rollers are often employed when spreading powders with limited flowability, as arises in powders comprising fine particle sizes or high surface energy materials.  Here, we study roller-based powder spreading for powder bed AM using the unique combination of a purpose-built powder spreading testbed with a proven method for X-ray mapping of powder layer depth. We focus on the density and uniformity of nominally $100$~\textmu m thick layers of roller-spread Ti-6Al-4V and Al-10Si-Mg powders.  Our results indicate that when rotation is too rapid, roller-applied shear force impedes the creation of dense and uniform layers from powders of high innate flowability, or where inertial forces driven by particle density dominate cohesive forces.  Roller counter-rotation augments the uniformity of cohesive powder layers, primarily though reducing the influence of particle clusters in the flowing powder, which are otherwise shown to cause deep, trench-like streaks.  Companion discrete element method (DEM) simulations further contextualize the experiments through isolation of the effects of cohesion on layer attributes.  Results suggest that roller motion parameters could apply a strategic level of additional shear force to the flowing powder, thereby mitigating the clumping behavior characteristic of highly cohesive feedstocks while maintaining high layer uniformity.
\end{abstract}

\begin{keyword}
additive manufacturing \sep powder \sep spreading \sep X-ray \sep quality \sep discrete element method
\end{keyword}

\end{frontmatter}

\section{Introduction}

Roller-based spreading mechanisms are commonly used in powder bed additive manufacturing (AM) processes, particularly when use of fine or highly cohesive powders conveys desirable process attributes and/or component properties~\cite{Ziaee2019, Avrampos2022}.  Broadly, powder bed AM functions by alternatively spreading thin layers of powder that are locally fused, e.g., by binder, sintering, or melting, to ultimately fabricate monolithic components one cross section at a time.  In these processes, layer thickness determines the vertical resolution of the printing process, which is the primary consideration in selecting this parameter, in addition to secondary considerations that potentially include process stability, build rate, post-process sintering conditions, and component mechanical performance~\cite{Ziaee2019, Nagarajan2019, Mindt2016, Vock2019, Brandt2017, Sutton2016, Brika2020, Heinl2020, Mussatto2021,Lee2015, Qiu2015OnMelting, Escano2018}.  However, powders that are large as compared to the layer height cause sparse deposition~\cite{Spierings2011, Mindt2016, Penny2021} and even jamming against the recoating implement~\cite{Nan2018}.  It follows that fine powders are desirable for these manufacturing processes, especially when binder jetting components destined for sintering, as to achieve accurate and fully dense components~\cite{Ziaee2019, Nagarajan2019, Mindt2016, Vock2019, Brandt2017, Sutton2016, Brika2020, Heinl2020, Mussatto2021}.

However, when fine particles are used in powder bed AM, the low particle mass increases the relative effect of cohesive forces, causing spontaneous clumping in the spreading flow. Therefore, fine powders may also be prone to spreading into sparse and highly variable layers~\cite{Escano2018, Meier2019CriticalManufacturing, Kiani2020, Muniz2018}.  Additional attributes impairing powder flowability and uniform layer formation include alloy composition, density, shape, oxidation, and contamination~\cite{Vock2019,Brika2020, Avrampos2022, Mussatto2021, Parteli2016, Snow2019, Wang2020}. In turn, this variability can manifest as component defects, including porosity, rough surface finish, and poor component accuracy~\cite{Ziaee2019, Lee2015, Qiu2015OnMelting, Mindt2016, Brika2020}.  Thus, enabling the manufacturing of high quality powder bed AM components lies at the intersection of discerning the powder attributes governing flowability in context, and imparting favorable spreading boundary conditions through selection of spreading strategy, or collectively the choice of the spreading tool(s) and motion parameters applied thereto.
 
A number of legacy powder characterization techniques exist, and have enjoyed a long history of quantifying the flowability of metal powders in practical contexts.  Among the simplest is the static angle of repose, wherein powder is permitted to flow through a standardized funnel and the slope of the resulting conical pile is measured.  Higher angles are positively correlated with decreasing powder flowability; a standardized funnel geometry designed by Hall as applied in ASTM standard B213~\cite{ASTM2013_ReposeAngle} is commonly used to measure this parameter.  Other approaches include the Hausner Ratio~\cite{Hausner1967} and Carr Index~\cite{Carr1965} that assess powder flowability through comparing bulk density to tap density.  While these methods are common and easy to perform, they lack specificity either to a targeted powder attribute or, on the other extent of the problem, to the fluctuating boundary conditions inherent to spreading in powder bed AM~\cite{Spierings2016, Vock2019, Kiani2020, Avrampos2022}.  Therefore, these simple methods, on their own, are poorly predictive of powder spreadability and layer attributes.

Accordingly, much experimental work centers on quantification of powder flow under conditions representative of AM. For example, Budding and Vaneker~\cite{Budding2013} use a purpose-built mechanized cup for extracting as-spread powder specimens of known volume to study binder jetting using a plaster powder (also see~\cite{Jacob2016, Wischeropp2019, Brika2020} for similar approaches).  They find that using a counter-rotating roller results in more dense layers than a rigid ('doctor') blade, and further that large diameter rollers give higher density layers than smaller diameter rollers.  This work additionally shows that forward rotation can be an effective means of increasing layer density via compaction (also see~\cite{Shanjani2008, Rishmawi2018}).  Parallel work by Meyer and coworkers~\cite{Meyer2017}, considering PA6 and PA12 nylon powders used in selective laser sintering (SLS), more closely examines the effect of surface speed in roller spreading by holding translational speed constant at 127 mm/s and varying roller rotational speed.  Significant decreases in packing density are observed with increasing surface speed, or sum of translational and roller tangential speed, from $153$ to $233$~mm/s, and further increasing surface speed $335$ to $607$~mm/s results in highly unpredictable spreading and more sparse deposition as compared to blade-based layers used as a control.  The authors conclude that there are three spreading regimes demarcated by roller surface speed: insufficient roller rotational speed causing high forces, compaction, and powder layer cracking; an optimal range with high density and high uniformity; and excessive rpm which throws powder forward and thus lowers packing density.  

Dispensing of cohesive powders, i.e., metering the amount needed to subsequently spread a layer, is also challenging. Hopper mechanisms are often used to pre-deposit highly cohesive powders onto the build area prior to roller-based leveling~\cite{Pruitt1991, Ziaee2019}.  Jimenez et al.~\cite{Jimenez2019} demonstrate that the density of $20-40$~\textmu m alumina powder layers deposited in this fashion depends strongly on the amount of powder deposited by the hopper by using an Archimedes approach to measuring specimen density.  Finally, our prior work led by Oropeza investigates roller-spreading of nominally $20$~\textmu m and $40$~\textmu m alumina powders, where the comparatively lower inertial forces of the finer powder manifest as lower flowability (angle of repose of $50.3^\circ$ versus $42.5^\circ$, respectively)~\cite{Oropeza2022}.  Divergent behavior is shown, whereby roller counter rotation ($300$~RPM) improves deposition of fine alumina powder ($20$~\textmu m nominal diameter) by more than a factor of two, yet impedes dense layer formation with powder having a larger ($40$~\textmu m nominal diameter) and more flowable size distribution.

Studies incorporating simulations have provided additional insights to the influence of spreading parameters in isolation.  For instance, Wang and coworkers, performing simulations based upon spreading of Ni-based alloy powders, conclude that knowledge of the coefficient of sliding friction and Hamaker constant (thus encoding the magnitude of van der Waals forces) are critical to capturing powder spreading kinematics~\cite{Wang2020}.  Increasing the Hamaker constant by two orders of magnitude is observed to reduce deposition by approximately 10\% in blade-spread layers, with increased layer surface roughness; however, use of a counter-rotating roller increases velocity gradients in the powder pile and thereby enhances particle flow into a dense and uniform layer.  Counterintuitively, a minor benefit of cohesion is noted; specifically, it helps keep fine particles uniformly distributed in the powder flow and thereby minimizes size segregation.  Follow-up work by the same group assesses simulated layers created using six different spreading tool geometries, using 70 and 90~\textmu m thick layers of 25.9 to 52.7~\textmu m (D$_{10}$ to D$_{90}$) nickel alloy powder as test cases~\cite{Wang2021}.  A blunt ($1$~mm radius) rounded blade is associated with high deposition and, critically, generates exceptional forces both along the downward and transverse (spreading) directions.  Mean forces with a counter-rotating roller are lower, as well as average deposition, although high force transients are still observed.  Finally, in knife-edge geometries, an inclined blade with a 45$^\circ$ bevel on the spreading edge is associated with modestly increased deposition as compared to 90$^\circ$ blade.  

Earlier work by Haeri~\cite{Haeri2017, Haeri2017_2} considers the effect of spreading tool as related to the aspect ratio of non-spherical powders.  Packing fraction is observed to monotonically decrease with increasing particle aspect ratio for blade-based spreading, as also seen experimentally by Brika and coworkers~\cite{Brika2020}, in contrast to roller-spread layers which achieve maximum density for particles with an aspect ratio of approximately 1.5.  This change in behavior arises from the roller preferentially aligning the major axis of the particles with the spreading direction and thereby achieving ordered and dense packing of the particles.  Thus, roller-based spreading may be favorable for high aspect ratio (e.g., substantially oblate) powders, although a rigid blade with an optimized super-elliptic profile may produce layers of nearly equivalent quality.  Research by Parteli considers the effect of roller translation speed, spanning a range of 20 to 180 mm/s, when spreading PA-12 nylon, and concludes that surface roughness of the powder layer increases quadratically as a function thereof.  Joint experiments and simulations illustrating a similar trend with a 316~SS powder are reported by Chen, and additionally show packing density to fall in a similar manner~\cite{Chen2020}.

Finally, we note that existing studies, including studies centered on blade-based spreading techniques, predominantly assess layer non-uniformity at process-relevant resolution (e.g., approximately the laser spot size in laser powder bed AM) via optical measurement of surface topology~\cite{ Craeghs2011_2, Kleszczynski2012, Hendriks2019, Lee2020, Snow2019, Zhang2016, Chen2020, Heinl2020, TanPhuc2019AManufacturing, Mussatto2021}.  However, this is a poor proxy for variation in the actual volume of powder deposited because the sub-surface configuration or packing density of powder layers is non-constant, in part due to a spatially evolving size distribution~\cite{Ali2018OnProcesses, TanPhuc2019AManufacturing}, particle-bridging effects~\cite{Chen2017}, and powder compaction at obstructions presented by local changes in boundary conditions~\cite{Penny2021}.  Notable exceptions, in addition to Oropeza et. al as mentioned above, include Escano and coworkers~\cite{Escano2018} who use a side-on synchrotron X-ray imaging technique to observe powder clustering dynamics in the spreading of stainless steel powders, and Beitz et al.~\cite{Beitz2019} who use micro-CT to study packing of oblate powders in blade-spread layers of PA12 nylon.

\begin{figure*}[htb]
\begin{center}
	{
	\includegraphics[trim = {1in 3.4in 1.1in 3.4in}, clip, scale=1, keepaspectratio=true]{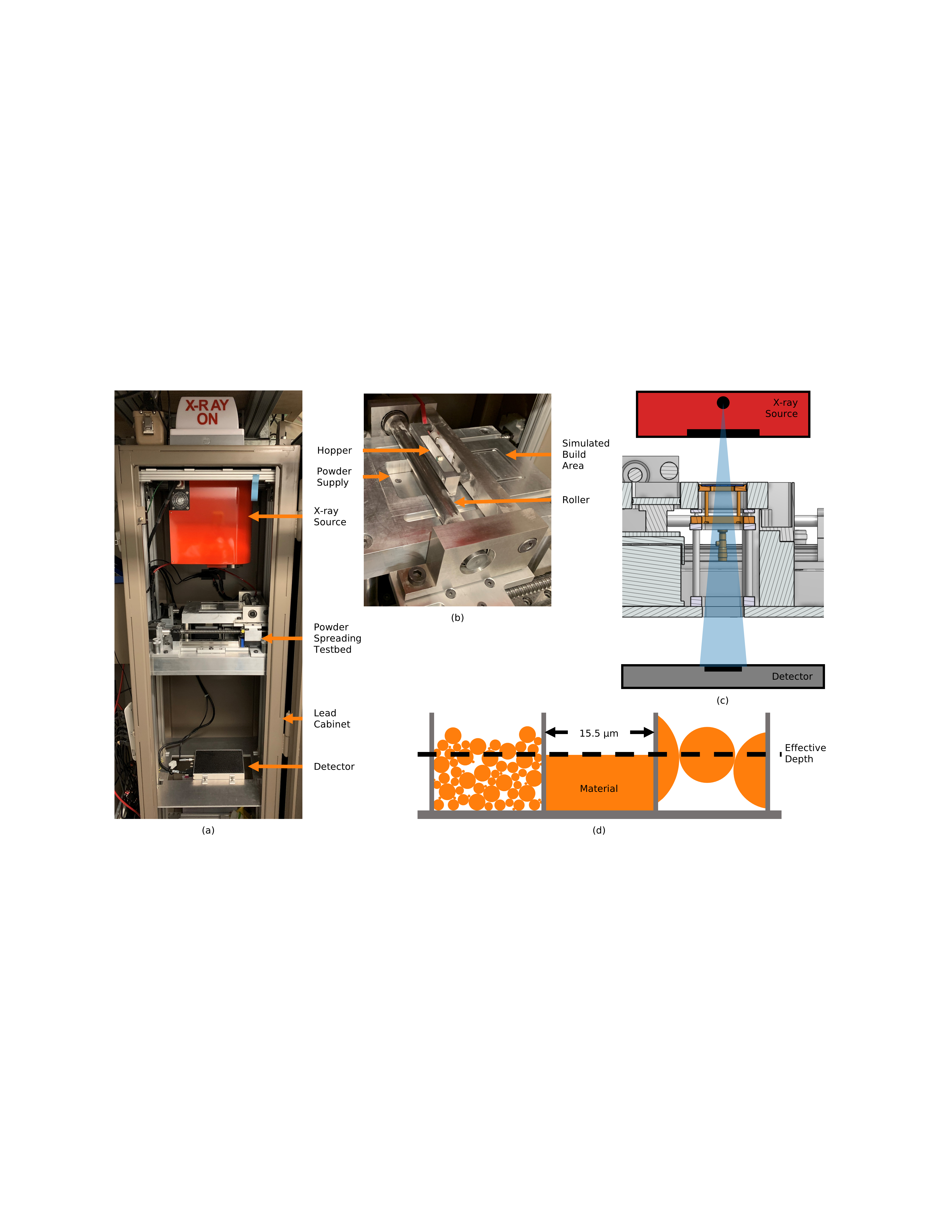}
	}
\end{center}
\vspace{-11pt}
\caption{X-ray interrogation of roller-spread powder layers. (a) X-ray imaging apparatus with the powder spreading testbed installed. (b) Close-up image of the testbed roller, powder supply, powder hopper, and build area.   (c) Schematic view of the X-ray beam path and simulated build plate.  (d) Graphical definition of powder layer effective depth, illustrating three equivalent material configurations.}
\label{fig:IntroFig}
\end{figure*}

Therefore, in the case of highly cohesive powders used in AM, there remains a critical need to quantify powder spreadability and layer non-uniformity, and to be able to resolve material deposition at process-relevant scales.  To this end, we apply the combination of a mechanized powder spreading testbed and X-ray mapping of powder layer effective depth, shown in Fig.~\ref{fig:IntroFig}, to the task of resolving deposition patterns of metal powders.  Powder layers are spread using a $20$~mm diameter roller with a range of motion parameters.  The effective depth of the powder layer, or thickness of material if the powder were to be fully consolidated without spatial redistribution (see graphical definition in Fig.~\ref{fig:IntroFig}d), is then mapped with the X-ray imaging system shown in Fig.~\ref{fig:IntroFig}a at $15.5$~\textmu m spatial resolution, enabling particle-scale layer non-uniformites to be resolved.  For example, Figs.~\ref{fig:ResultsStats}b and~\ref{fig:ResultsStats}c visually show marked decrease in layer uniformity when the traverse speed of the roller is increased from $5$ to $50$~mm/s.  Roller spreading parameters are systematically evaluated using $15-45$~\textmu m Ti-6Al-4V powder as a baseline case, including statistical and power spectral density analysis of layer uniformity.  Additionally, the effects of increased particle cohesion are experimentally studied with a $20-63$~\textmu m Al-10Si-Mg powder, as well as parallel DEM simulations that enable isolated study of feedstock cohesive forces and particle density.

\label{sec:Intro}
\section{Methods}
\label{sec:Methods}

\subsection{Mechanized Powder Spreading}

Powder layers are created using the mechanized spreading testbed developed and qualified by Oropeza et al. in~\cite{Oropeza2021}, configured for these experiments as shown Figs.~\ref{fig:IntroFig}b and~\ref{fig:IntroFig}c.  In combination with the ballscrew-actuated traverse stage, the system is capable of roller speeds spanning $0$ to $300$~RPM and traverse speeds of up to $200$~mm/s.  The instrument further comprises a build area designed to enable transmission X-ray measurements, as is schematically illustrated in Fig.~\ref{fig:IntroFig}c, where the X-ray beam path is shown in blue.  To achieve a nominally $100$~\textmu m thick powder layer, two micrometer heads (one is visible in the cross-sectional view of Fig.~\ref{fig:IntroFig}c) are used to position the $2$~mm thick, 6061 aluminum build platform $100$~\textmu m below the roller using a gap gauge (Starrett No. 467M).  The build platform height is adjusted until a slight drag is felt as the gauge is slid along the roller, and the final settings are validated by ensuring the gauge slides freely at $90$~\textmu m and cannot be inserted into the gap at $110$~\textmu m.  For reasons that become apparent in the following results, runout is a critical parameter in the implementation of roller-based powder spreading.  Qualification of this parameter via laser line scanner in~\cite{Oropeza2021} gives $15$~\textmu m total indicated runout over the central $10$~mm of the roller.

To create a powder layer, a volume of powder is first expressed from a supply piston (to the left of the representative build area depicted in Fig.~\ref{fig:IntroFig}b), and then is collected as the roller sweeps over it.  This powder is then deposited both over the build area and surrounding mechanical structure, including filling the gaps therebetween.  As the X-ray procedure described below requires thorough cleaning after imaging a first layer prior to collecting a baseline image for a second trial, all powder outside of the supply piston is removed.  Thus, to ensure that sufficient powder is spread as to fill the gaps and prevent a short-feed, 4 times the nominal amount of powder is expressed by the supply piston as would theoretically be required to cover the $75 \times 75$~mm build area of the instrument at the desired layer thickness.  Extremely cohesive powders (e.g. $0-25$~\textmu m Ti-6Al-4V) cannot be spread by this method, as the powder is too cohesive to flow between the roller and build area.  However, such powders can be directly deposited over the build area with a vibrating hopper, shown in Fig.~\ref{fig:IntroFig}b, that is affixed to the traverse stage ahead of the roller.  As configured here, this apparatus deposits powder at approximately $0.3$~g/s as it is moved across the build area.

\subsection{X-ray Measurement of Effective Depth}
Measurement of powder layer effective depth is performed via X-ray transmission, using a custom-built apparatus described by Penny et al.~\cite{Penny2021}.  Figure~\ref{fig:IntroFig}a identifies the critical components of the apparatus, comprising a Hamamatsu L12161-07 X-ray source placed at the top of a lead cabinet (Hopewell Designs) that generates a beam passing downwards, through the build platform of the mechanized spreading testbed, and finally to a CMOS flat panel detector equipped with a cesium iodide scintillator (Varex Imaging 1207 NDT).  These instruments are positioned on an aluminum frame, enabling adjustable geometric magnification; as disposed here, the detector resolves a $23.8 \times 13.3$~mm$^{2}$ area centered on the build platform with $15.5$~\textmu m/px spatial (lateral) resolution.  X-ray imaging is performed with the source set to a potential of $50$~keV and current of $200$~\textmu A, enabling use of the small ($5$~\textmu m) focus mode of the source (i.e., resolution is limited by the combination of pixel pitch and geometric magnification, and not by blurring from the finite emission volume of the source).  In turn, an integration time of $18000$~ms provides signals of approximately $80$\% of the dynamic range of the detector.

Transmission of the powder layer is assessed as the ratio of an image with a powder layer present to a previously recorded baseline image of the build platform without powder.  As uncertainty on transmission is dominated by shot noise, each image comprises a sum of individual frames to improve the signal-to-noise ratio.  Specifically, $52$ frames are summed for measurements of Ti-6Al-4V, and $520$ frames are summed for Al-10Si-Mg due to its lower X-ray stopping power, hence lower contrast. These values achieve $1$~\textmu m uncertainty in the effective depth measured by each detector pixel.  To keep the duration of experiments feasible, in view of the stark difference in experiment duration, $5$ model layers are studied in experimental combinations including Ti-6Al-4V and one layer is imaged when Al-10Si-Mg powder is used.  X-ray measurements are automated to synchronously control the source and detector, average individual frames, and periodically measure and correct for detector dark current.

Transmission measurements are interpreted as effective depth using our forward radiation transport model \cite{Penny2021}, comprising: evaluation of the source emission spectrum after Birch and Marshall~\cite{Birch1979}; calculating the transmission through objects in the beam path, including the powder layer and build platform, using Lambert's law~\cite{Lambert1760} and spectral attenuation coefficients from NIST~\cite{Saloman1988X-ray92}; and detector scintillator light yield (from~\cite{Holl1988}) and gain deduced from observation of noise statistics.  With this model, transmission is computed as a function of powder layer depth, including accurate representation of non-linearity induced by beam hardening.  As applied here, this function is densely sampled, then numerically inverted to recover effective depth from observed transmission.

\subsection{Powders}

\begin{figure}[ht]
\begin{center}
	{
	\includegraphics[trim = {3in 4.4in 3in 4.4in}, clip, scale=1, keepaspectratio=true]{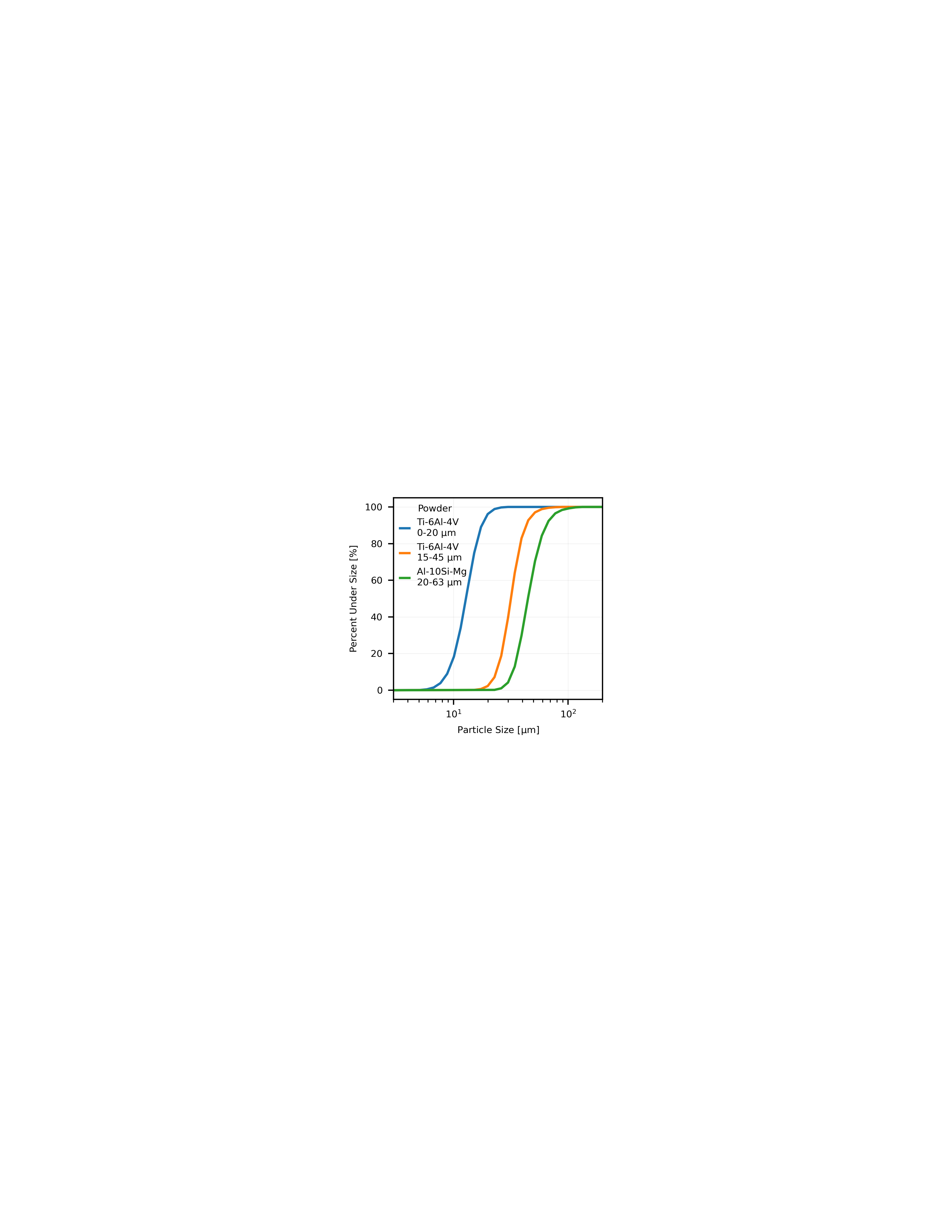}
	}
\end{center}
\vspace{-11pt}
\caption{Cumulative size distributions for the three powders studied herein.}
\label{fig:MethodsPowderSize}
\end{figure}

The baseline powder for the experiments conducted herein is a nominally $15-45$~\textmu m Ti-6Al-4V powder that, in view of our prior study with a manual powder spreading approach, features moderate flowability as compared to the range of powders commonly used in metal powder bed AM.  We contrast this powder with $20-63$~\textmu m Al-10Si-Mg; despite having a modestly coarser size distribution, it is considerably more cohesive and less flowable as evidenced by its higher angle of repose (AoR) and Hausner ratio.  This arises from its lower mass density, and therefore higher relative magnitude of cohesive forces as compared to inertial forces when flowing.  Finally, one experimental case is studied using $0-20$~\textmu m Ti-6Al-4V, which is an extremely cohesive powder.  In fact, this powder is so cohesive as to preclude angle of repose measurement with a Hall-specified (\cite{ASTM2013_ReposeAngle}) funnel, though, due to different sensitivities to particle properties~\cite{Spierings2016}, has a lower Hausner ratio than the $20-63$~\textmu m Al-10Si-Mg powder. Table~\ref{table:Method_Powders} compares the nominal powder sizes to their actual measurements via laser diffraction (Horiba LA960), as well as AoR measured via Hall flowmeter funnel~\cite{ASTM2013_ReposeAngle}.   Figure~\ref{fig:MethodsPowderSize} shows substantially log-normal distributions of the powders via cumulative size data.

Knowledge of the exact powder composition is critical to ensuring the precision of the X-ray measurement technique.  Accordingly, the composition reported for the specific powder lots on the material certification certificates is used in modeling powder layer attenuation in the aforementioned radiation transport model, as opposed to the nominal proportion of alloy elements.

\begin{table*}[htb]
\centering
\footnotesize
\caption{Tabulated powder properties. * the $0-25$~\textmu m Ti-6Al-4V powder is sufficiently cohesive to preclude standardized measurement of AoR, as it does not flow through a funnel fabricated to Hall flowmeter specifications.}
\label{table:Method_Powders}
\begin{tabular}{@{}*8c@{}}
  
  \toprule[1.5pt]
  \multicolumn{1}{c}{\head{Material}} &
  \multicolumn{1}{c}{\head{Nominal Size [\textmu m]}}&
  \multicolumn{1}{c}{\head{D10 [\textmu m]}} &
  \multicolumn{1}{c}{\head{D50 [\textmu m]}}&
  \multicolumn{1}{c}{\head{D90 [\textmu m]}}&
  \multicolumn{1}{c}{\head{Angle of Repose [$^\circ$]}}&
  \multicolumn{1}{c}{\head{Hausner Ratio}} &
  \multicolumn{1}{c}{\head{Supplier}}\\
  
  \cmidrule{1-8}
 
    Ti-6Al-4V & 15-45 & 23.4 & 31.5 & 43.4 & 35.6 & 1.12 &AP\&C (GE) \\
    Ti-6Al-4V & 0-20 & 9.3 & 14.0 & 19.1 & * & 1.27 & AP\&C (GE) \\
    Al-10Si-Mg & 20-63 & 32.7 & 44.7 & 64.9 & 38.2 & 1.44 & IMR Metal Powders\\
 
  \bottomrule[1.5pt]
\end{tabular}
\end{table*}

\subsection{Power Spectral Density Analysis}
In addition to the elementary statistical methods applied to effective depth maps of powder layers presented below, we make use of spatial frequency domain techniques to more fully understand the length scales of variance within powder layers. Described in this context in~\cite{Penny2021}, our analysis centers on first projecting the X-ray measurements of effective depth into the spatial frequency domain by performing a 2 dimensional Fourier transform.  The resulting complex amplitude data are squared, giving the power spectral density (PSD) as a function of spatial frequency, or, equivalently, variance as a function of inverse dimension.  As normalized here, integrating (summing) the entire PSD gives a value equal to the statistical variance of the powder layer.  More deeply, however, fluctuations of a given characteristic dimension are mapped to a radius about the origin of the PSD plot.  Direction is also preserved.  As an example germane to the data at hand, variation from a sinusoidal pattern of effective depth exactly aligned with the spreading direction maps to a point\footnote{Two points are associated with this pattern mathematically, as the Fourier transform, and therefore the PSD, of a real-valued function is symmetric about the origin in frequency space.  This distinction is omitted here for clarity; however, the integrations described herein necessarily respect the equivalency of positive and negative spatial frequencies in this case.} in the PSD. The specific location lies along the axis passing through origin of the PSD that is aligned with the direction of spreading, and, again, at a radius corresponding to its spatial frequency.

Thus, to assist in interpreting these data, the PSD is fully integrated as a function of direction to evaluate the total variance associated with a size scale, regardless of orientation.  In certain cases, we alternatively restrict integration of the PSD to a specific range of directions, namely those within $\pm 15^\circ$ of the spreading direction and the direction normal thereto.  Accordingly, layer nonuniformity may be ascribed to these directions, to assist in determining root cause.  We do not apply this technique to the simulated layers; their small spatial extent, while sufficient to assess average and overall variance, provides an inadequate number of measurements to support PSD analysis.

\subsection{DEM Modeling}

Experimental work is complemented by discrete element method (DEM) powder spreading simulations  implemented in the parallel multi-physics research code BACI~\cite{Baci}. We refer the reader to \cite{Meier2019ModelingSimulations,Meier2019CriticalManufacturing, Meier2021GAMM, Penny2021} for additional descriptions of this DEM model framework and calibration procedures. The powders are modeled with spherical discrete elements, wherein the size distribution is fitted to experimental laser diffraction measurements of the powders used in corresponding experiments. More specifically, values for the $10^{th}$-percentile, median and $90^{th}$-percentile are fitted to a lognormal distribution for this purpose.  From a physical perspective, the DEM model incorporates interactions between particles as well as between particles and structural elements (i.e., the substrate and spreading tool), that are considered in the equations of motion of each particle. Structural elements, however, are fully displacement-controlled, i.e., they are modeled as fully rigid and deflections are not considered in this study. Normal forces consist of repulsive contact forces modeled via a spring-dashpot penalty model as well as cohesive van-der-Waals forces, characterized by their pull-off force. Tangential (frictional) forces are modeled via a spring-dashpot approach coupled to the normal force through Coulomb's law.  Additionally, rolling resistance is considered in the equation of angular momentum.

For discretization of the roller geometry, a tessellated model of a $10$~mm diameter roller (idealized, without runout) is represented by $125$ linear segments around the theoretical roller perimeter. This size is a trade-off between the size of $20$~mm diameter in the experimental setup and computational costs for simulating a realistically sized powder ensemble relative to the roller. A sensitivity study shows no significant impact of this smaller diameter on the spreading results, along with the moderate number of $42,000$ particles in total for the given size distribution. In order to simulate comparable circumferential velocities, the rotational velocity of the roller is scaled with the diameter ratio, i.e., a $20$~mm roller at $250$~RPM in the experiment is modeled with a $10$~mm roller at $500$~RPM . As applied in our previous studies \cite{Meier2019ModelingSimulations,Meier2019CriticalManufacturing, Meier2021GAMM, Penny2021}, the dimension of the simulation perpendicular to the spreading direction is $1$~mm with periodic boundary conditions, and the powder bed has a length of $12$~mm.  However, only a $7$~mm long segment located at the center of the powder bed is used in the analysis to mitigate potential edge effects at the beginning or the end of the powder bed. Among other metrics that may be extracted, effective depth is evaluated in analogy to the experimental X-ray technique with a spatial resolution of $15.5$~\textmu m.  Following the procedure described in~\cite{Penny2021}, the intersection length of particles with vertical rays is determined on a fine grid and then down-sampled (averaged) to match the spatial resolution of the experiment.

A pseudo-material modeling approach is proposed to study the effects of particle surface energy and density.  Surface energy, which determines cohesive forces, is systematically varied.  Thus, for each parameter set consisting of traverse velocity and rotational velocity, surface energy values ranging from $0.02$ to $2.56$~mJ/m$^2$ (non-cohesive to very cohesive) are simulated.  We justify this range in view of our prior study, which suggests that a surface energy in the range of $0.04-0.08$~mJ/m$^2$ best captures the behavior of the $15-45$~\textmu m Ti-6Al-4V powder with good flowability.  Conversely, powders with surface energies $>1$~mJ/m$^2$, at the other end of the chosen range, are extremely cohesive.  It is assumed that the same surface energy used to describe particle-particle interactions is also applied to particle-boundary interactions.  Additionally, in order to better understand the isolated impact of different material densities, the same pseudo-material approach is applied to the Al-10Si-Mg powder. Here, all parameters, including those defining the surface energy sweep, are are chosen identically to Ti-6Al-4V except for the particle density.

\section{Results}
\label{sec:Results}

\subsection{Effect of Roller Motion Parameters on Effective Depth}

\begin{figure*}[ht]
\begin{center}
	{
	\includegraphics[trim = {1.4in 3in 1.4in 2.9in}, clip, scale=1, keepaspectratio=true]{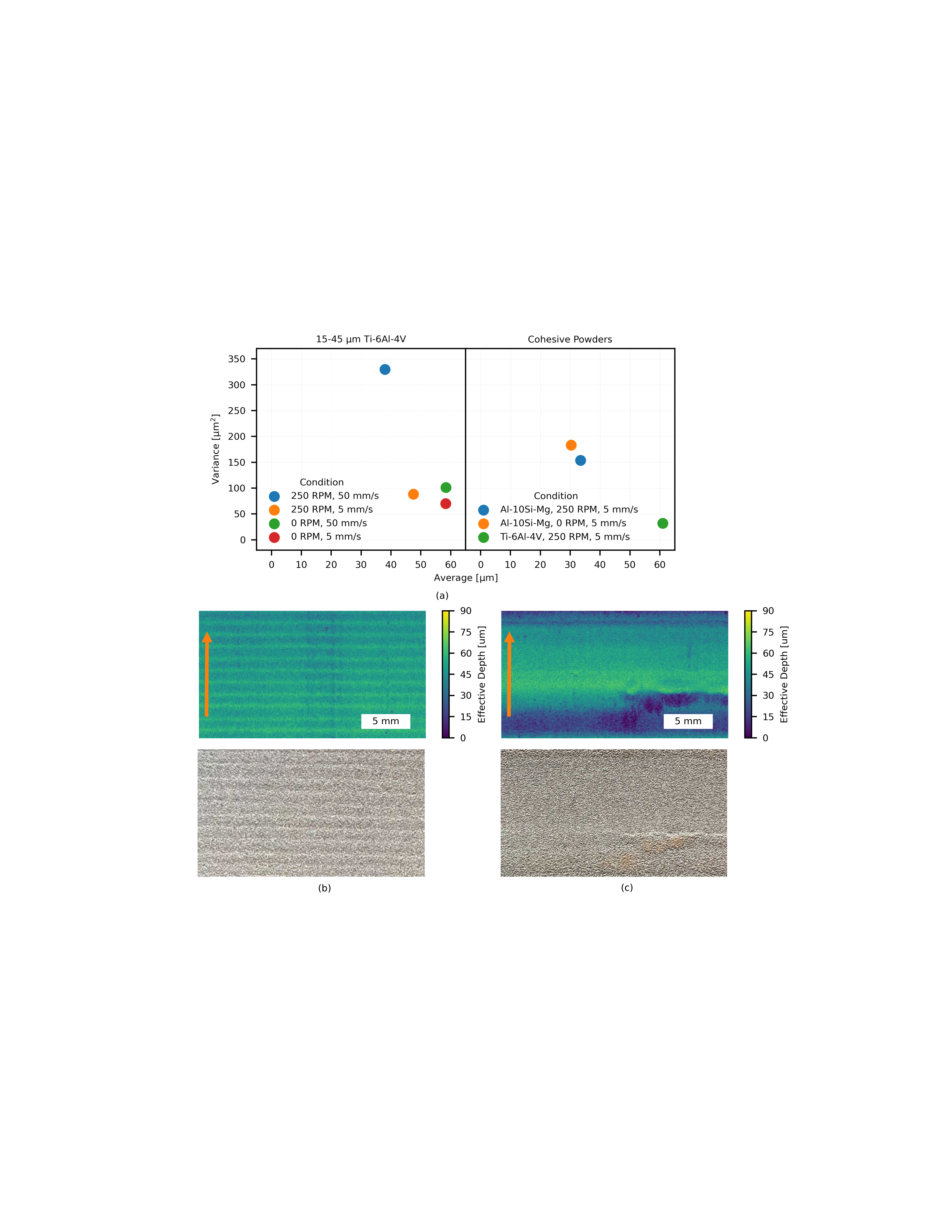}
	}
\end{center}
\vspace{-11pt}
\caption{Summary of experimental layer characteristics.  (a) Statistical data corresponding to layers of $15-45$~\textmu m Ti-6Al-4V, left, and more cohesive powders, right.  (b-c) Comparison of nominally $100$~\textmu m layers of Ti-6Al-4V spread with $250$~RPM roller rotation at traverse speeds of $5$ and $50$~mm/s, respectively, via processed X-ray (top) and optical (bottom) images. The orange arrows indicate the spreading direction.}
\label{fig:ResultsStats}
\end{figure*}

We present results for the influence of roller counter-rotation and traverse speed, beginning with the simple layer statistics in Fig.~\ref{fig:ResultsStats}a, in which the variance in effective depth is plotted versus the average depth for the experimental powder layers.  The left panel presents the $15-45$~\textmu m Ti-6Al-4V results, and one may immediately conclude that the combination of roller rotation and high traverse speed ($250$~RPM and $50$~mm/s, respectively) only serves to create layers of undesirably high variance.  Lowering the traverse speed to $5$~mm/s reduces layer variance by more than a factor of three and increases average effective depth by nearly $10$~\textmu m.  Layers fabricated without roller rotation show slightly increased deposition, achieving an average effective depth of $\approx 60$~\textmu m, and again a slower traverse speed results in lower layer variance.  Compared to the cohesive (Al-10Si-Mg) powder layers, quantified in the right panel, the opposite trend is observed with respect to roller rotation.  Namely, roller rotation decreases layer variance from $183$ to $154$~\textmu m$^2$, while simultaneously increasing average effective depth from $30.3$ to $33.5$~\textmu m.

\begin{figure*}[hbt]
\begin{center}
	{
	\includegraphics[trim = {1.1in 4.3in 1.1in 4.3in}, clip, scale=1, keepaspectratio=true]{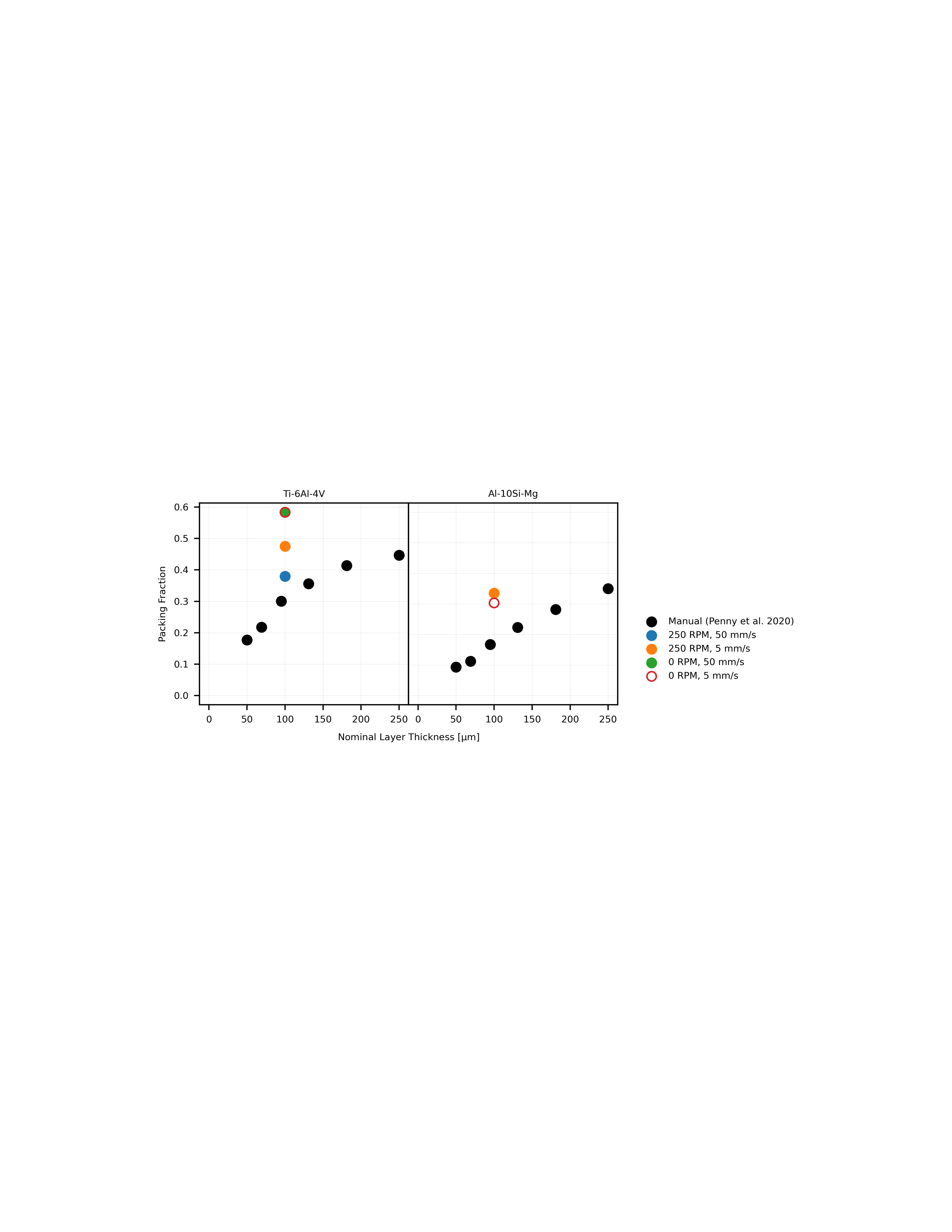}
	}
\end{center}
\vspace{-11pt}
\caption{Experimental packing fraction of roller-spread powder layers, compared to manually spread layers in a range of nominal thicknesses from~\cite{Penny2021}.}
\label{fig:SupplementManual}
\end{figure*}

Leveraging Fig.~\ref{fig:SupplementManual}, we compare these results to our prior work with these powders in~\cite{Penny2021}.  Therein, these powders are manually spread at low ($\approx 5$~mm/s) into precision-etched silicon templates to generate layers with a range nominal thicknesses, using a $1/8$~in. thick machinist's flat as a blade.  Results from this study indicate that the packing fraction exponentially approaches an asymptotic value, somewhat analogous to a pour packing fraction, with increasing layer thickness.  The packing fractions achieved with the cylindrical blade geometry in our current results lie well above our manual spreading trend, suggesting that this blade shape achieves a more dense packing of powder particles.  In Ti-6Al-4V, the $250$~RPM, $5$~mm/s case very nearly achieves the asymptotic value of $0.5$ from the manual spreading experiments, indicating that little room remains for further densification of the powder layer.  The packing fraction in the $0$~RPM cases exceeds both this asymptotic value and the tap packing fraction of $0.56$ determined from manufacturer's data, suggesting that recoating forces are sufficient to cause an extreme degree of compaction.  The asymptotic packing fraction of Al-10Si-Mg under manual spreading is $0.59$ (c.f. $0.73$ tap packing fraction calculated from manufacturer's data).  Again, the roller geometry achieves a more dense particle configuration.  The packing fractions still lie well below the asymptote, however, showing that the particle configuration is not optimally dense and may be considerably improved via spreading implement design and motion parameters.

\begin{figure*}[ht]
\begin{center}
	{
	\includegraphics[trim = {1.2in 3in 0.6in 0.4in}, clip, scale=1, keepaspectratio=true]{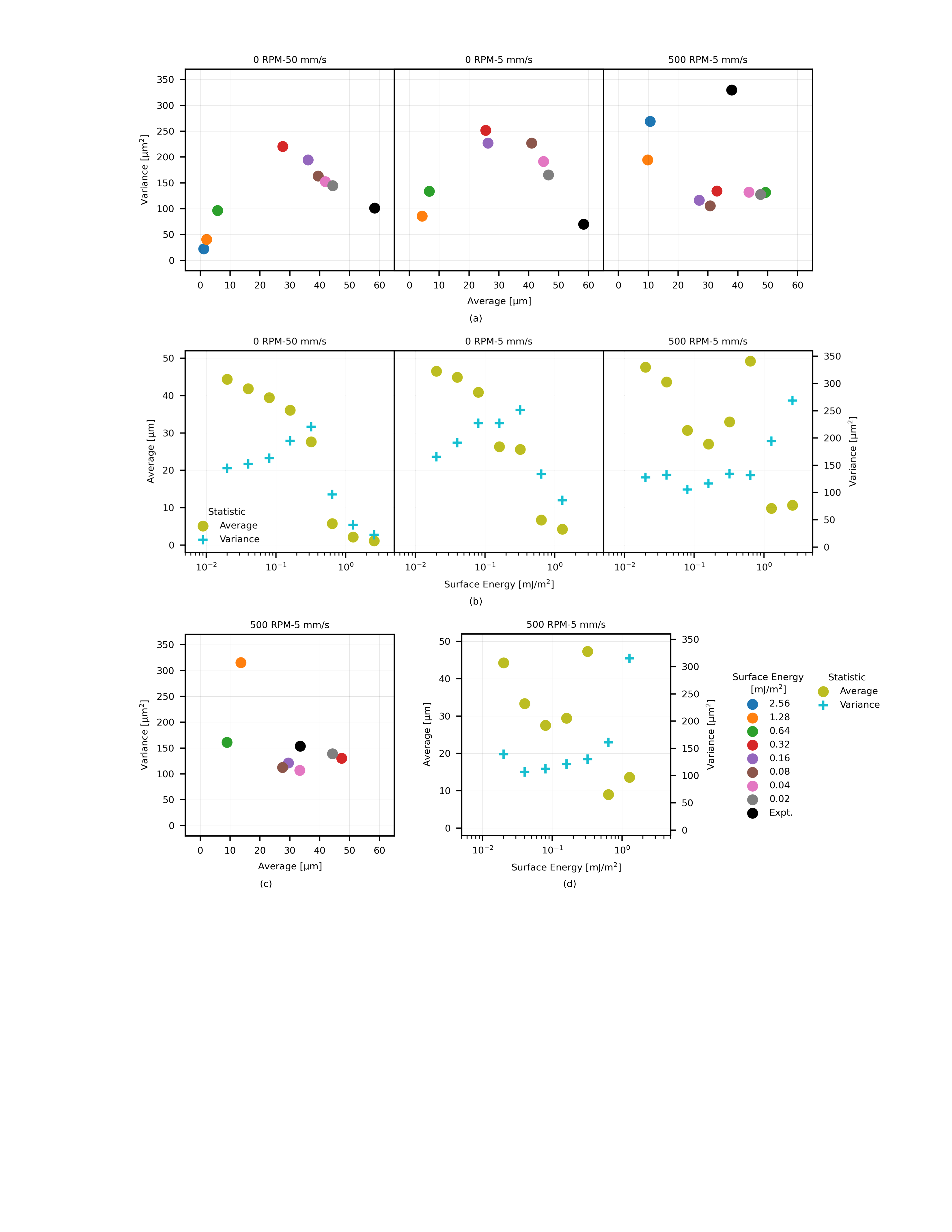}
	}
\end{center}
\vspace{-11pt}
\caption{Statistics of simulated layers, nominally $100$~\textmu m thick, created with a selection of spreading implements.  (a) Pseudo-material parameters selected to replicate the particle size and density of $15-45$~\textmu m Ti-6Al-4V for a range of surface energies.  (c) Pseudo-material particle density adjusted to replicate Al-10Si-Mg.  (b,d) Alternate representation of (a,c), respectively, wherein statistics are plotted against surface energy.}
\label{fig:ResultsSimStats}
\end{figure*}

\begin{figure*}[ht]
\begin{center}
	{
	\includegraphics[trim = {1.0in 3.8in 0.85in 4in}, clip, scale=1, keepaspectratio=true]{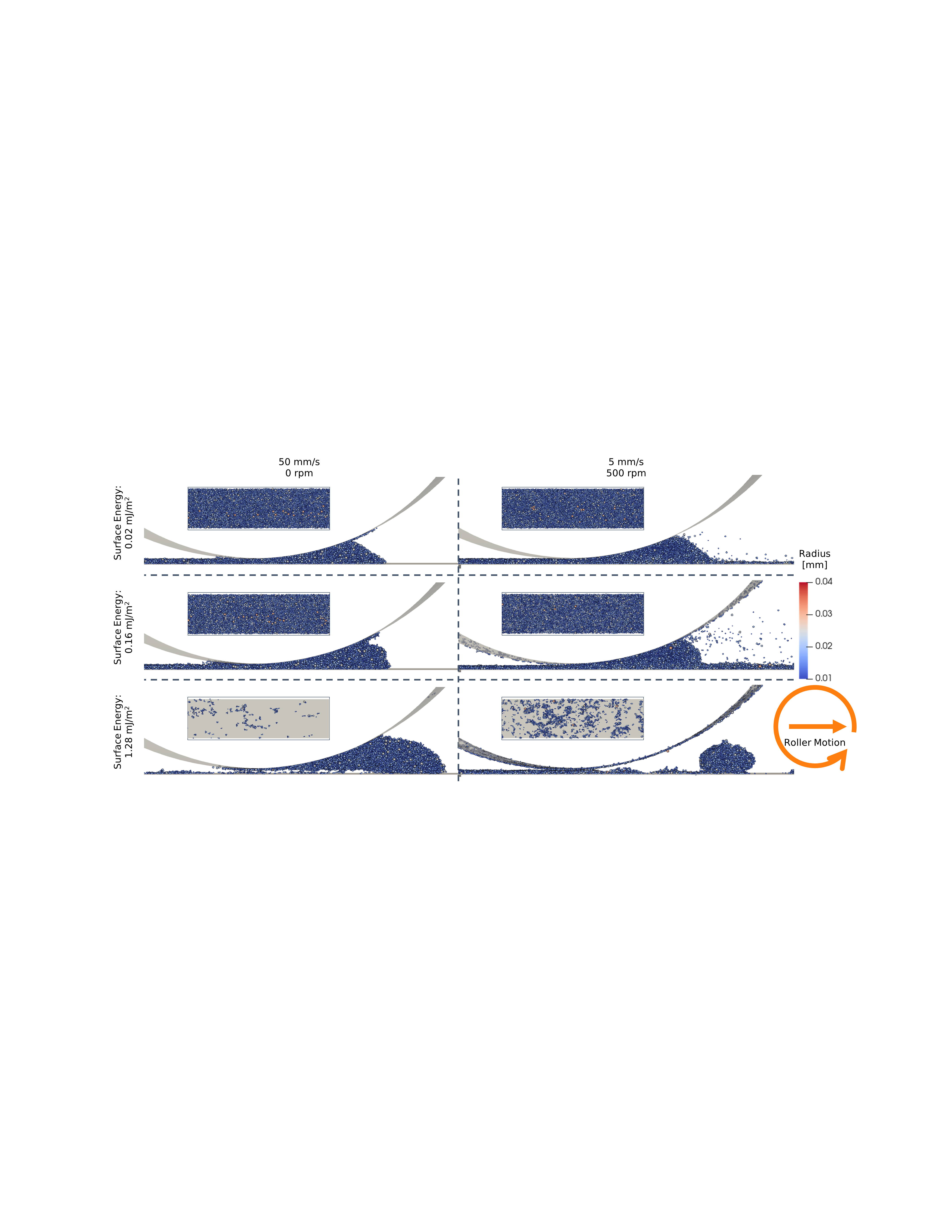}
	}
\end{center}
\vspace{-11pt}
\caption{Images extracted from simulations demonstrating differences in flow behavior for a selection of spreading parameters and powder surface energies.  Inserts show a $1 \times 3.5$~mm$^2$ portion of the layer, top-down.  Material size distribution and density corresponds to $15-45$~\textmu m Ti-6Al-4V.}
\label{fig:ResultsSimulations}
\end{figure*}

Figure~\ref{fig:ResultsSimStats}a compares the experimental results for roller-based spreading of $15-45$~\textmu m Ti-6Al-4V with DEM simulations, performed with a deliberately altered range of surface energies (see Fig.~\ref{fig:ResultsSimStats}b for an alternate representation of these data).  In the left panel, describing the simulated case of $0$~RPM and $50$~mm/s, we observe that average layer depth falls and variance increases as powder surface energy is increased from $0.02$~mJ/m$^2$ (i.e., powder flowability decreases) and deposition becomes more irregular. This parallels a qualitative reading of Fig.~\ref{fig:ResultsSimulations}, in which the corresponding motion parameters induce clumping and visually less smooth powder flow in the $0.16$~mJ/m$^2$ case than the $0.02$~mJ/m$^2$ case.  At a value between $0.32$ and $0.64$~mJ/m$^2$, the powder becomes so extremely cohesive as to be largely pushed forward as one large clump by the roller, and therefore little powder is deposited in the layer.  In other words, powder flow closely resembles the $1.28$~mJ/m$^2$ panel of Fig.~\ref{fig:ResultsSimulations}, wherein the powder adheres so severely to itself, the build plate, and roller as to cause a bridging effect.  In this high surface energy regime, variance decreases with increasing cohesiveness, as the build area is largely devoid of powder particles. The indicated experimental result is at the low-cohesion end of this trend, with slightly lower variance and higher deposition than expected by the DEM simulation.  Reducing the traverse speed to $5$~mm/s typically results in higher layer depth, as seen in the center panels of Figs.~\ref{fig:ResultsSimStats}a and~\ref{fig:ResultsSimStats}b, especially in the case of highly cohesive powders.  Layer variance in the high surface energy regime is notably higher than in the $50$~mm/s case and is explained by the fact that powder is deposited, i.e., variance is no longer artificially low because a great number of regions are bare.  

Results from simulations at the counter-rotating roller speed of $500$~RPM and traverse speed of $5$~mm/s are shown in the right panel of Fig.~\ref{fig:ResultsSimStats}a, and the more complex trend is again best understood with the corresponding plot in Fig.~\ref{fig:ResultsSimStats}b.  At surface energy below $0.16$~mJ/m$^2$, the trend resembles the non-rotating case at at the same spreading speed and average deposition falls monotonically with increasing surface energy.  However, beyond this point counter-rotation causes the trend in effective depth to reverse, coming to a sharp peak at approximately $0.64$~mJ/m$^2$ before rapidly falling off.  Thus, the additional shear provided by the roller can assist in depositing powders over a targeted range of increased surface energies.  Finally, we note that the simulation is in agreement with the experimental results, where counter-rotation modestly reduces deposition at surface energies comparable to the physical Ti-6Al-4V powder ($0.04$ to $0.08$~mJ/m$^2$).

Roller counter-rotation improves layer uniformity (i.e., lower variance) for powders of surface energy below $0.64$~mJ/m$^2$.  This may be qualitatively understood via the simulation images in Fig.~\ref{fig:ResultsSimulations}.  Specifically, powder flow in the $500$~RPM, $5$~mm/s, $0.16$~mJ/m$^2$ case more closely resembles the bulk powder flow of the low cohesion ($0.02$~mJ/m$^2$) case than the high cohesion case, whereas the tortured shape of the avalanche slope in the non-rotating case ($0$~RPM, $50$~mm/s, $0.16$~mJ/m$^2$) reveals irregular powder flow.  At and above $0.64$~mJ/m$^2$ the variance of effective depth is higher than in the non-rotating case; yet, without rotation the layers are highly sparse and thus, while having low variance, are not suitable for AM. This is again made plain in Fig.~\ref{fig:ResultsSimulations}, where spreading an extremely cohesive powder (surface energy of $1.28$~mJ/m$^2$) with roller rotation results in a more dense and uniform layer, despite disturbance from large clumps that are periodically propelled across the build platform by the advancing rotating roller, as compared to the non-rotating case.  

Next, we compare the average and variance of effective depth in simulations predictive of spreading $15-45$~\textmu m Al-10Si-Mg powder, as to decouple the effects of surface energy and density, as presented in Figs.~\ref{fig:ResultsSimStats}c and~\ref{fig:ResultsSimStats}d.  The results are similar to the equivalent Ti-6Al-4V case, namely decreasing deposition with increasing surface energy, then reversing to a narrow peak, and falling sharply as the powder becomes too cohesive to effectively deposit.  However, the Al-10Si-Mg curve is shifted slightly to the left, indicating that flowability of the less-dense powder is more sensitive to surface energy (e.g., the peak occurs at approximately $0.16$~mJ/m$^2$).  Again noting the size difference between the simulated and experimental powders, the experimental result closely matches the expected deposition and variance for intermediately cohesive ($0.04$ to $0.16$~mJ/m$^2$) powder simulations.  At this range of surface energies the trend again matches the experimental results, albeit opposite that expected for the Ti-6Al-4V powder.  Namely, the simulation and experiment agree that adding counter-rotation increases deposition for this more cohesive powder.  Last, we note that the slightly higher variance observed in the experiment is largely explained by the washboard pattern on the powder layers (e.g., as seen in Figs.~\ref{fig:ResultsStats}b and~\ref{fig:ResultsStats}d) caused by the slight runout of the roller in the spreading apparatus.   

\subsubsection{Cumulative Distribution Functions}

\begin{figure*}[ht]
\begin{center}
	{
	\includegraphics[trim = {2in 3.4in 2in 3.4in}, clip, scale=1, keepaspectratio=true]{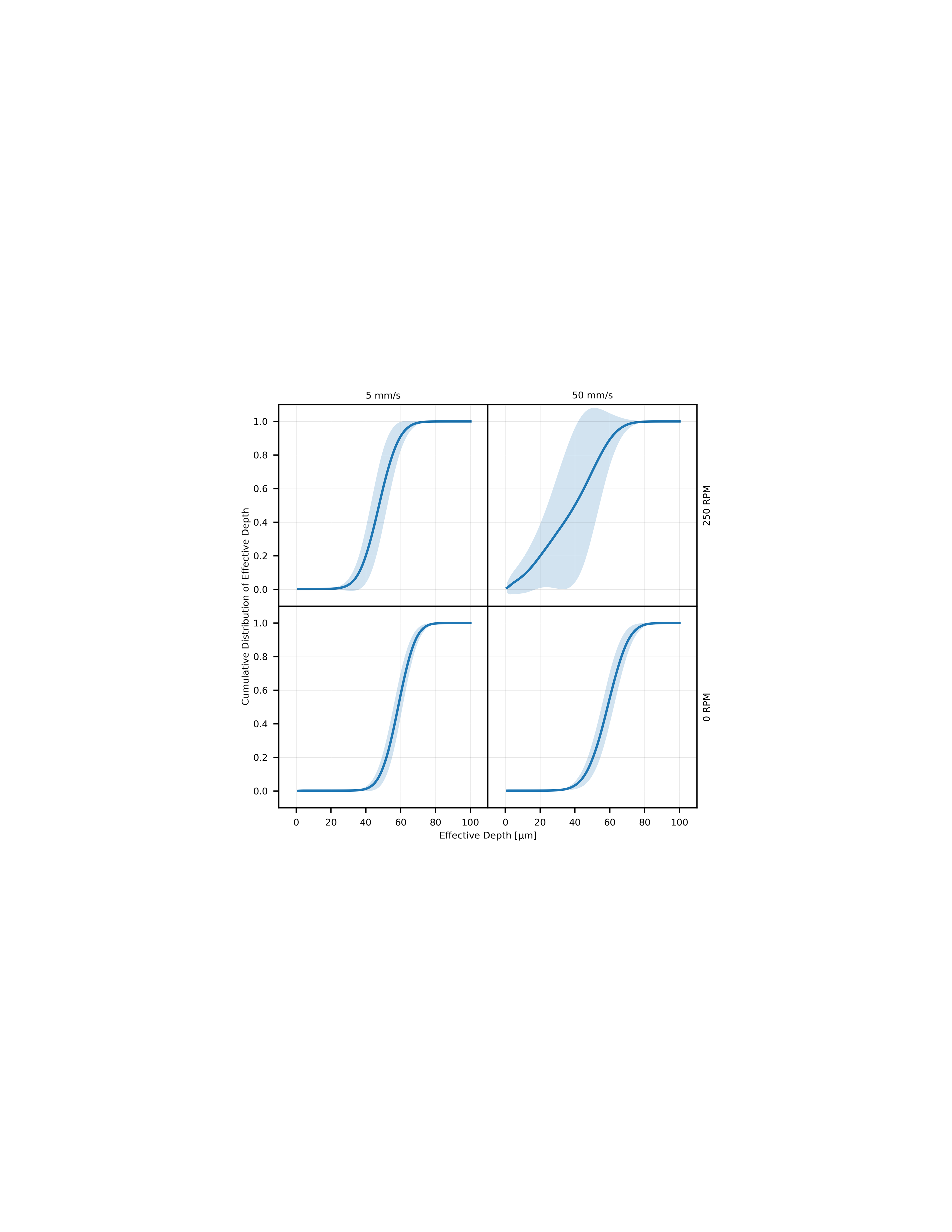}
	}
\end{center}
\vspace{-11pt}
\caption{Cumulative distribution functions of experimental $15-45$~\textmu m Ti-6Al-4V layers, including the average and $\pm 3 \sigma$ range.}
\label{fig:ResultsTiHist}
\end{figure*}

\begin{figure}[ht]
\begin{center}
	{
	\includegraphics[trim = {2.65in 3.1in 2.7in 3in}, clip, scale=1, keepaspectratio=true]{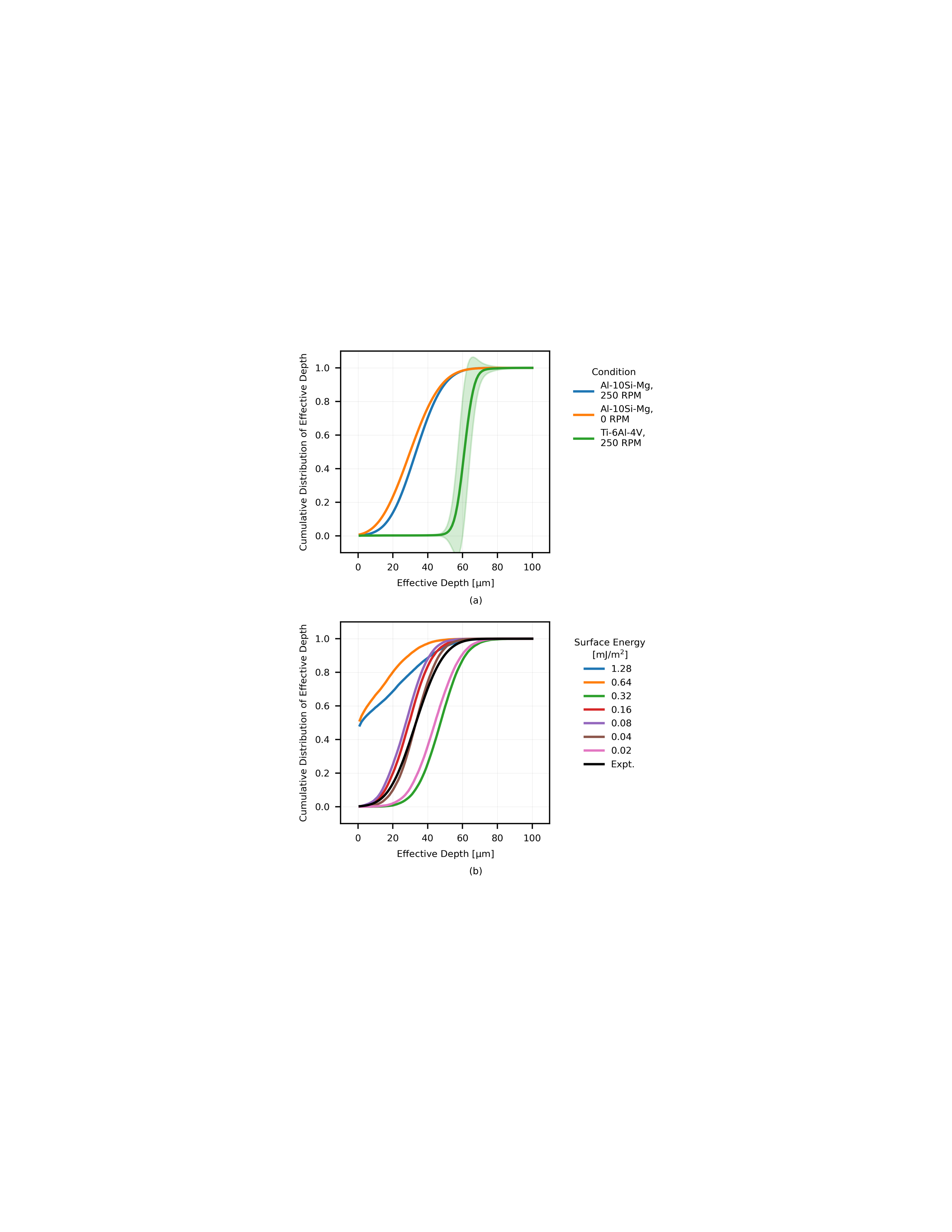}
	}
\end{center}
\vspace{-11pt}
\caption{Cumulative distribution functions of layers fabricated with cohesive powders. (a) Experimental data comparing the effect of roller rotation in spreading layers of Al-10Si-Mg, as well as provisional data on redistributing $0-25$~\textmu m Ti-6-Al-4V after hopper deposition.  (b) DEM simulation results corresponding to $15-45$~\textmu m Al-10Si-Mg powder spread with a roller speed of $500$~RPM and traverse speed of $5$~mm/s for a selection of surface energies.  Note: The corresponding experimental curve is shown in black.}
\label{fig:ResultsAlHist}
\end{figure}

We reinforce this understanding by constructing effective depth cumulative distribution functions (CDFs) from the effective depth maps of experimentally spread powder layers in Figs.~\ref{fig:ResultsTiHist} and~\ref{fig:ResultsAlHist}a.  Figure~\ref{fig:ResultsTiHist} comprises the average depth distribution of the 5 $15-45$~\textmu m Ti-6Al-4V experimental layers spread in each condition in bold, and the shaded region further denotes the $\pm 3 \sigma$ range observed over the 5 model layers.  Again, the $50$~mm/s, $250$~RPM trial is the clear standout in terms of poor layer quality, showing a non-zero probability of completely bare regions, poor consistency, and a non-gaussian distribution shape.  Both the layer-to-layer consistency and mean distribution width (variance) improve when the traverse speed is reduced to $5$~mm/s.  Layers spread without roller rotation are also very consistent across the trial layers, and are similar in distribution shape to the $5$~mm/s, $250$~RPM layers, albeit with higher average deposition.  This distribution shape is also observed in the experimental Al-10Si-Mg histograms in Fig.~\ref{fig:ResultsAlHist}a.  Again, however, the trend is reversed for the cohesive powder, where roller rotation is seen to primarily enhance deposition by suppressing the low-effective-depth tail of the distribution.  We support this observation statistically, noting that the distribution skew falls from $0.284$ to $0.193$ with the benefit of roller counter-rotation.

\begin{figure*}[ht]
\begin{center}
	{
	\includegraphics[trim = {2in 3.4in 2in 3.4in}, clip, scale=1, keepaspectratio=true]{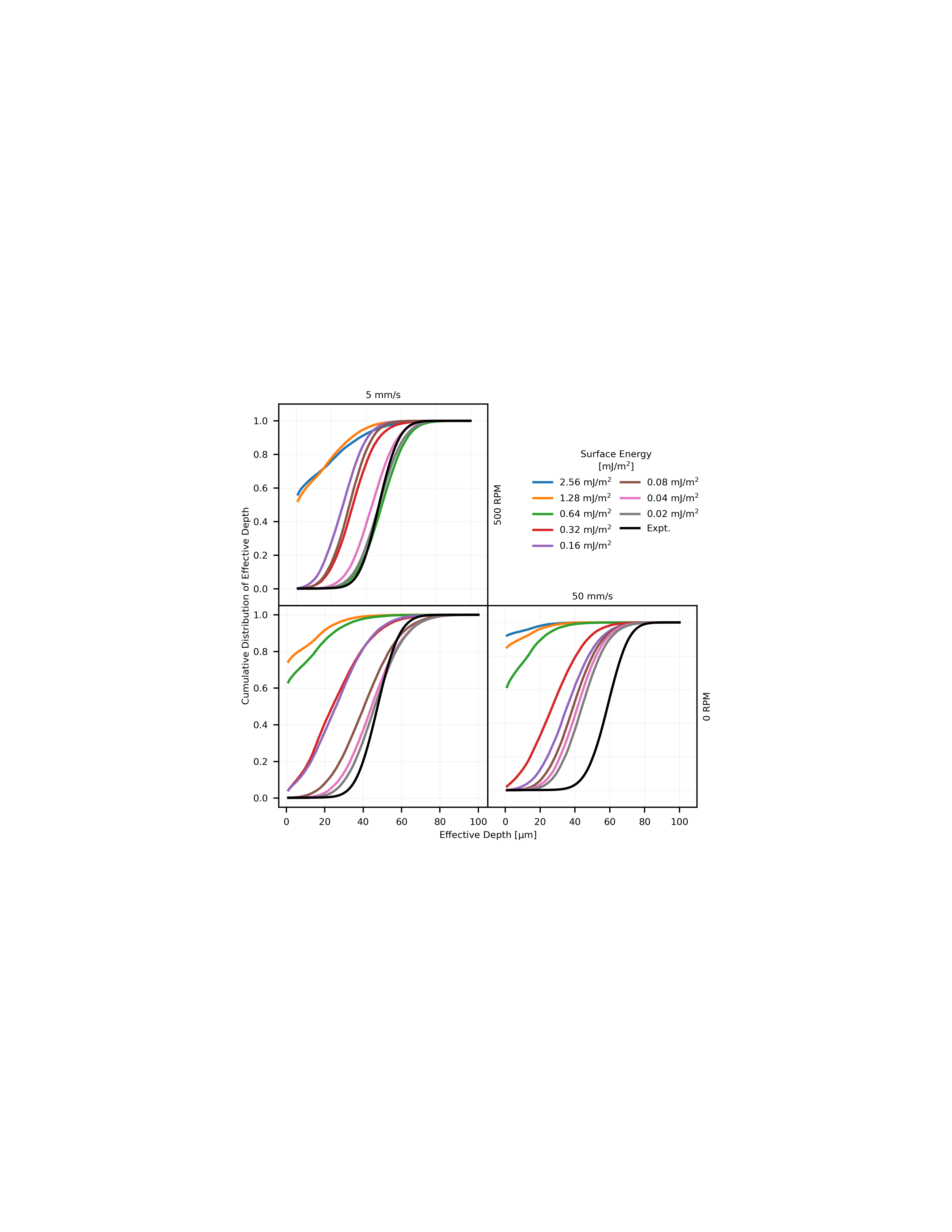}
	}
\end{center}
\vspace{-11pt}
\caption{Cumulative distribution functions of DEM-simulated Ti-6Al-4V layers, varying presence of roller rotation, traverse speed, and surface energy. Note: To facilitate comparison to Fig.~\ref{fig:ResultsTiHist}, the average experimental curve is plotted in black.}
\label{fig:ResultsTiSimsHist}
\end{figure*}

Figures~\ref{fig:ResultsTiSimsHist} and~\ref{fig:ResultsAlHist}b present the corresponding simulated distributions for spreading of Ti-6Al-4V and Al-10Si-Mg, respectively.  These results clarify how, at low traverse speed, roller counter-rotation enables improved results in spreading cohesive powders.  Specifically, we observe in Fig.~\ref{fig:ResultsTiSimsHist} that percent coverage (i.e., the fraction of build platform area covered by powder at any nonzero effective depth) improves for all of the most cohesive powers (surface energies spanning $0.32$ to $2.56$~mJ/m$^2$) simulated.  In comparison to the experiments, we again note particularly tight agreement in the low-traverse-speed, counter-rotating case, where the distribution shape and position closely matches simulation results corresponding to highly flowable powders, and under-prediction of powder deposition in the non-rotating simulations.  Similar findings are noted for the Al-10Si-Mg simulation CDFs in Fig.~\ref{fig:ResultsAlHist}b, where again extremely cohesive powders (i.e. with surfaces energies of approximately $0.64$~mJ/m$^2$ and greater for this material density) are shown to spread poorly with high likelihood of bare regions.  Layer density and quality, or low variance and low probability of bare regions, peaks at surface energies around $0.32$~mJ/m$^2$, and the shape and location of the experimental curve best matches the $0.04$~mJ/m$^2$ simulation. 

\subsubsection{Power Spectral Density Analysis}

\begin{figure*}[ht]
\begin{center}
	{
	\includegraphics[trim = {0.9in 2.7in 1.4in 3in}, clip, scale=1, keepaspectratio=true]{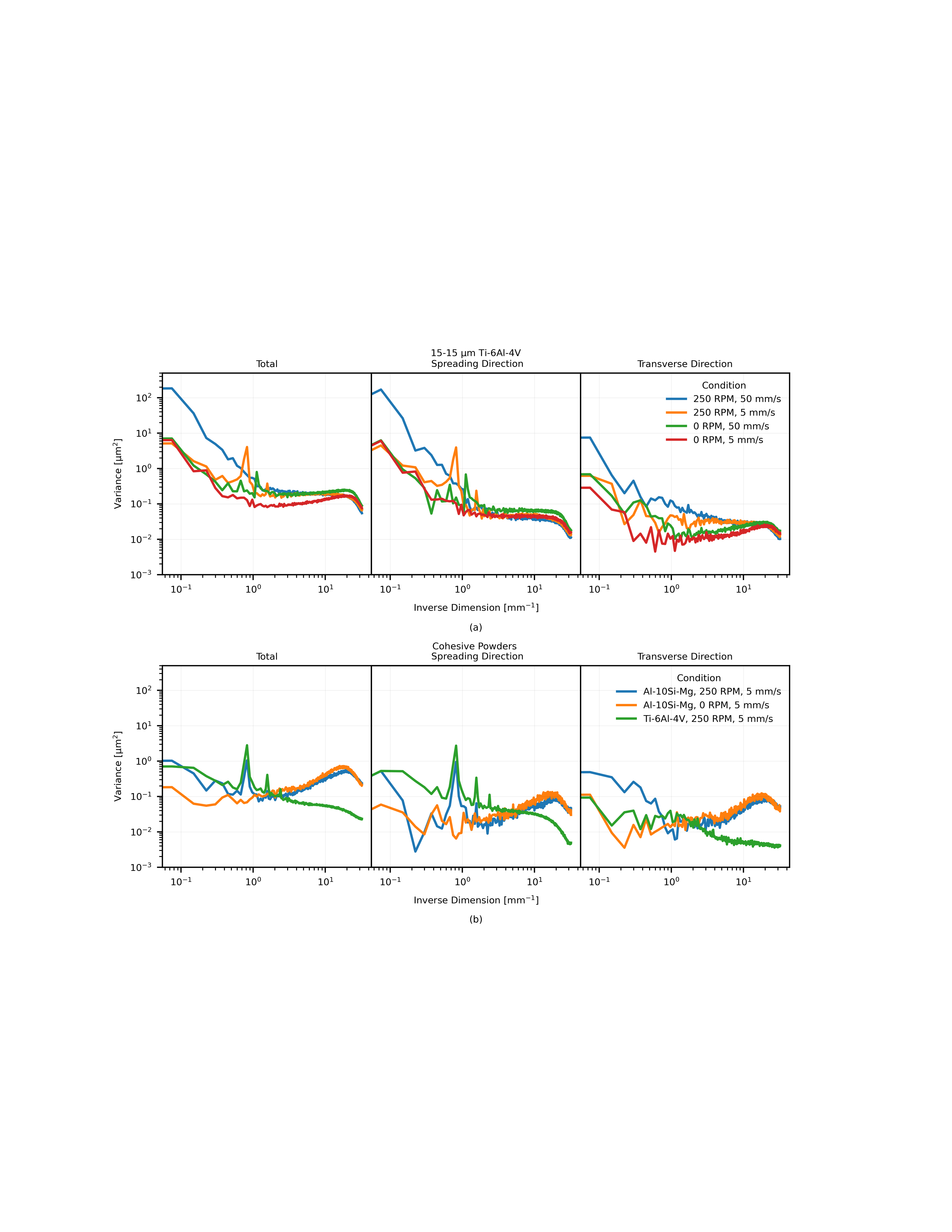}
	}
\end{center}
\vspace{-11pt}
\caption{Power spectral density analysis of powder layer variance based on X-ray transmission data.  (a) Comparison of spreading parameters in experimental layers of $15-45$~\textmu m Ti-6Al-4V. (b) PSD curves illustrating the effect of increased particle cohesion, by comparison to results for Al-10Si-Mg and leveling of hopper deposited $0-20$~\textmu m Ti-6Al-4V.}
\label{fig:ResultsPSD}
\end{figure*}

Via PSD analysis of the experimental data, we consider layer variance as a function of inverse dimension in Fig.~\ref{fig:ResultsPSD}.  Beginning with the $15-45$~\textmu m Ti-6Al-4V data in the left panel of Fig.~\ref{fig:ResultsPSD}a, the curve for 250~RPM and 50 mm/s again stands out due to its comparatively high variance. This, as previously described, is primarily due to the fluctuations in powder deposition visible at the millimeter scale or longer in Fig.~\ref{fig:ResultsStats}c.  Moreover, the layers are strongly anisotropic at these length scales, with more variance in the spreading direction due to the severe washboard pattern, expected at an inverse defect dimension of $0.083$~mm$^{-1}$ (or $\approx 12$~mm in conventional dimensions) by the combination of roller and traverse speeds, and dictated by the runout of the roller mechanism.  Decreasing the traverse speed to $5$~mm/s reduces the low frequency variance substantially and improves isotropy, aside from the peaks corresponding to the now $\approx 1$~mm long washboard pattern and its harmonics.  Based upon a rough approximation~\footnote{Here, we take the indicated runout of the roller as a $15$~\textmu m P-V, or $7.5$~\textmu m amplitude sinusoidal error that is transferred to the surface of the powder.  Further assuming a packing fraction of $50$\%, the sinusoidal disturbance in effective depth has an amplitude of $A = 3.25$~\textmu m.  The variance associated with such a pattern is $A^2/2$, or $\approx 5$~mm$^2$.}, the $15$~\textmu m roller runout is expected to manifest as a $5$~mm$^2$ increase in variance that closely matches the height of the spike in the $5$~mm/s case.  As the spike is much higher ($\approx100$~mm$^2$) in the $50$~mm/s case, we find that the high traverse speed amplifies the effect of roller runout.  Eliminating the roller rotation improves layer uniformity by eliminating these spikes in the spreading direction.  Further, this condition, with no rotation and slow traverse speed, features the lowest variance, especially at intermediate inverse dimensions.

Finally, we note two points concerning these curves in the high inverse dimension ($>20$~mm$^{-1}$) extreme, which may be expressed as $50$~\textmu m or a few mean particle diameters in conventional units.  First, at approximately this point for the layers spread without counter rotation, there lies a broad peak in variance in the transverse direction.  This arises from the narrow axis of streak-like defects, or regions of low deposition caused by interactions of the roller with powder clumps.  Naturally, these streaks also correspond to comparatively elevated variance in the spreading direction at lower inverse dimensions, thereby capturing the elongated aspect ratio of these defects.  This peak, or streaking behavior, is most pronounced in the lowest-shear ($0$~RPM, $5$~mm/s) case.  The curves from layers created with roller rotation, and thus under conditions of elevated shear in the powder pile, show no such peak, and instead variance smoothly rolls off at this point.  Second, for both low and high traverse speeds, total layer variance with rotation lies below the corresponding curve without rotation above this inverse dimension value.  Thus, we conclude that increasing applied shear improves layer uniformity at fine length scales by mitigating spontaneous clumping of particles in the flow, but the additionally turbulent powder flow reduces uniformity when considering larger, millimeter-scale areas.

This view is further supported though analysis of Al-10Si-Mg model layers, presented in Fig.~\ref{fig:ResultsPSD}b.  Clearly, adding roller rotation induces a spike in variance associated with the roller runout, but also increases variance at all inverse dimensions below $\approx 1$~mm$^{-1}$, as seen in the Ti-6Al-4V study.  Above this value, the hump associated with particle agglomeration is even more obvious, arising from the more cohesive nature of this powder.  However, roller rotation clearly causes a decrease in layer variance at high inverse dimension, with the difference peaking in the $10-20$~mm$^{-1}$ range.  In this case, the lower variance at high inverse dimension clearly outweighs the increase at longer length scales, resulting in more uniform layers in absolute terms by mitigating clumping.

\subsection{Roller Rotational Speed}

\begin{figure*}[htb]
\begin{center}
	{
	\includegraphics[trim = {0.85in 4.5in 1in 1.5in}, clip, scale=1, keepaspectratio=true]{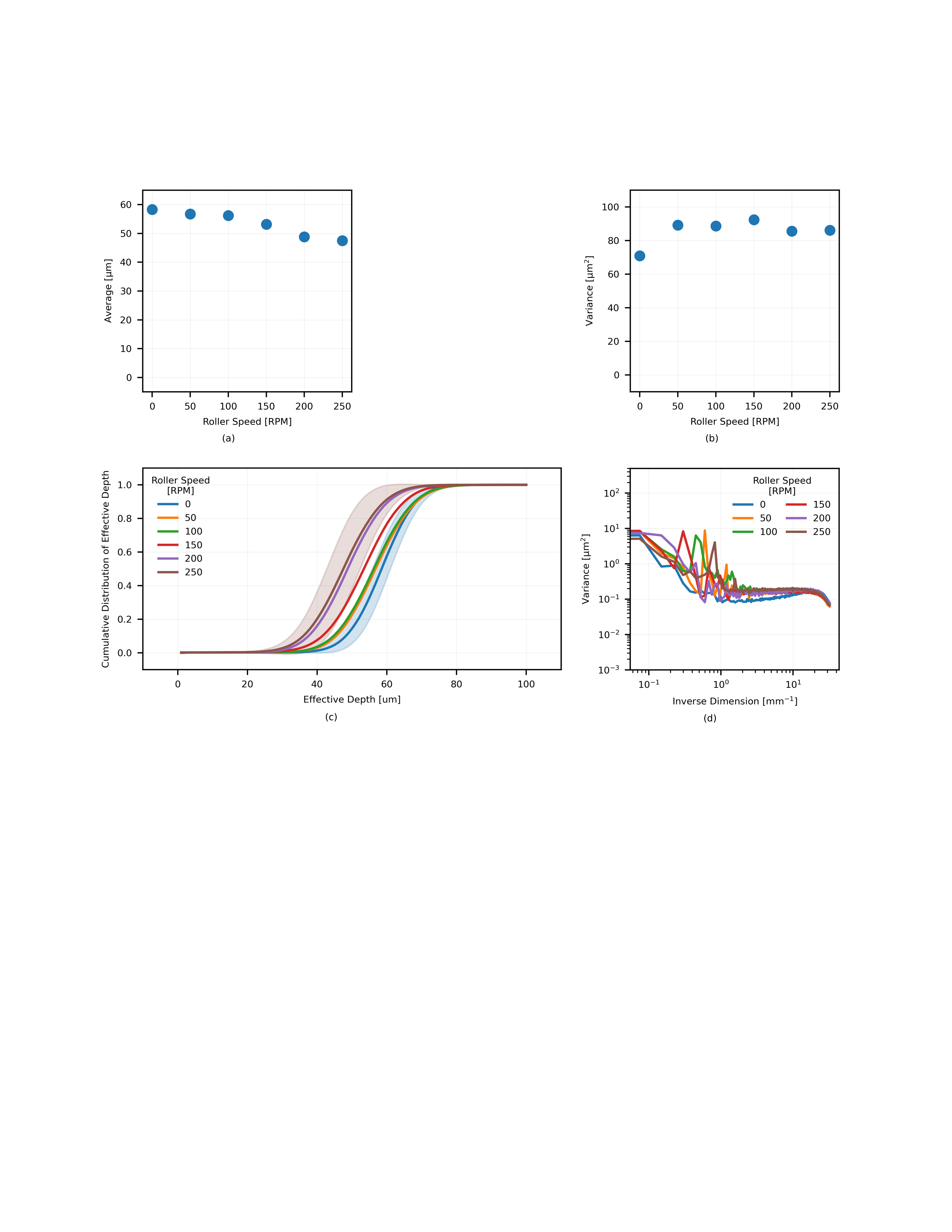}
	}
\end{center}
\vspace{-11pt}
\caption{Effect of roller counter-rotation speed on the effective depth of experimental powder layers.  (a-b) Comparison of layer statistics, illustrating decreasing layer density with increasing roller speed.  (c) Cumulative distribution function analysis showing similar distribution shape.  (d) PSD analysis exhibiting consistent layer variance outside of the peak associated with roller runout.}
\label{fig:ResultsRPM}
\end{figure*}

Figure~\ref{fig:ResultsRPM} presents an experimental parameter sweep to study the impact of roller rotational speed on the spreadability of the $15-45$~\textmu m Ti-6Al-4V powder, where the aforementioned $0$~RPM and $250$~RPM data are augmented with measurements of single layers spread at $50$, $100$, $150$, and $200$~RPM.  For this moderately flowable material, we observe that the densest, most uniform layer, and also with the best layer-to-layer variability, is created absent counter-rotation.  In cases with counter-rotation, layer variance is approximately constant at $\approx75$~\textmu m$^2$, and average deposition falls monotonically with increasing roller speed.  Finally, Fig.~\ref{fig:ResultsRPM}c shows no appreciable difference in the size scales associated with layer variance as a function of roller speed, outside of the conclusions related to roller runout described above.  We conclude that the effect of roller counter-rotation is negative for flowable powders, as the variance associated with roller runout outweighs the benefit of fracturing fine powder clumps that are already unlikely to affect powder flow in low-cohesion materials, and further reduces layer density.  However, we do note that increasing roller speed provides an effective means of increasing shear forces on powder particles and clusters thereof, which may be tailored to the flow characteristics of more cohesive powders.

\subsection{Redistribution of Extremely Cohesive Powders}

Finally, we consider layers fabricated from $0-20$~\textmu m Ti-6Al-4V powder, which is so extremely cohesive as to preclude flowability measurement via standard AoR methods.  This poor flowability precludes piston-feeding the powder for subsequent spreading.  Practically no deposition results for the range of spreading parameters (traverse and roller speeds) considered herein, and the powder is pushed across the build area in a singular large clump. However, we are able to study layers of this powder by pre-depositing it via hopper mechanism, and locally redistributing the powder using the roller to form a the final layer.  Figure~\ref{fig:ResultsStats}a illustrates that this method results in a dense layer of high effective depth and the lowest variance of the experimental layers, and Fig.~\ref{fig:ResultsAlHist}a further shows that the layer-to-layer consistency is similar to that of the coarser Ti-6Al-4V distribution.  PSD analysis in Fig.~\ref{fig:ResultsPSD}b shows that, as compared to piston-fed $15-45$~\textmu m Ti-6Al-4V, the hopper deposited $0-20$~\textmu m Ti-6Al-4V creates layers of comparable variance at intermediate inverse dimension, but notably lower variance at high inverse dimension (approximately above $3$~mm$^{-1}$).  The more uniform layers at high spatial frequency likely arise from the combination of fine particle size and increased fracturing of particle clusters from exceptionally high spreading forces.  We conclude, by separating the functions of bulk powder delivery from localized redistribution, that roller-spreading provides an effective avenue towards the use of finer powders in powder bed AM. This is consistent with established commercial approaches using a hopper mechanism, followed by one or more rollers, to spread thin layers of fine, highly cohesive powders in binder jetting AM systems.

\section{Conclusion}

By X-ray transmission mapping of powder layers deposited using a mechanized testbed, in combination with DEM simulation, we demonstrates how the effects of roller counter rotation and traverse speed interact with powder cohesion to strongly influence powder layer quality.  Roller counter rotation is shown to benefit spreading of cohesive powders, as the increased shear forces are sufficient to break apart powder clusters.  This benefit does not extend to all powders, however, as roller rotation is shown to increase layer non-uniformity, including resulting in periodic patterns in effective depth arising from roller runout.  Thus, an implement of static geometry (i.e., a blade) may be indicated for powders of high innate flowability, which do not require high applied forces to overcome inter-particle adhesion.  DEM simulations further suggest a threshold surface energy, relative to powder material density, beyond which powders become too cohesive to reliably spread.  However, this point may be extended to more cohesive powders by adding counter rotation.  Finally, roller mechanism runout is demonstrated to be a major determinant of layer uniformity, implying that roller spreading of fine powders in low layer thicknesses demands very high rotational precision of the spreading mechanism.

Upon this basis, future work should center on closely matching spreading parameters to the characteristics of specific powders in two directions.  First, our experiments and simulations both suggest that spreading parameters may be specifically tailored to achieve high density and low variance when spreading cohesive powders.  Thus, more densely sampling the range of powders, traverse speeds, and roller parameters (e.g., speed, diameter, material, and surface finish), is central to developing spreading guidelines for future feedstocks used in powder AM.  This is currently being investigated via the DEM simulation described herein, specifically considering forces arising from motion of the spreading and their effects on powder flow stability.  Second, we envision additional study of hopper-based powder delivery strategies for highly cohesive powders, in view of the high-quality layers provisionally achieved here.  Reliable metering of fine powders followed by spreading enables improved resolution and future use of exotic materials of high surface energy in AM.

\section{Acknowledgements}

Financial support at MIT was provided by: Honeywell Federal Manufacturing \& Technologies (FM\&T); a gift from Robert Bosch, LLC; a MathWorks MIT Mechanical Engineering Fellowship (to R.W.P.); a NASA Space Technology Research Fellowship (to D.O.); and a research grant from Lockheed Martin Corporation. We also thank Rachel Grodsky (Honeywell FM\&T) for performing laser diffraction particle sizing measurements. P.P., C.M., and W.W. acknowledge funding of this work by the Deutsche Forschungsgemeinschaft (DFG, German Research Foundation) within project 414180263.

\bibliographystyle{elsarticle-num}
\bibliography{refs.bib}

\beginsupplement
\FloatBarrier
\section{Comparison to Manual Spreading Experiments}

\FloatBarrier

\section{Computational Modeling Parameters}

\begin{table}[htb]
\centering
\caption{DEM model parameters}
\label{tab:DEMtarameters}
\renewcommand{\arraystretch}{0.7}
\begin{tabular}{ l l l } 
 \toprule
 Parameter & Value & Unit \\ 
 \midrule
 \multirow{2}{*}{Density}       &Ti-6Al-4V: 4430      &kg/m$^3$     \\ 
       &Al-10Si-Mg: 2670      &kg/m$^3$   \\ 
 Penalty parameter      &0.34      &N/m   \\ 
 Poisson's ratio     &0.342         &- \\
 Coefficient of friction     &0.4         &- \\
 Coefficient of rolling resistance     &0.07         &- \\
 Coefficient of restitution     &0.4         &- \\
 Surface energy     &varied: 0.02-2.56         &mJ/m$^2$ \\
 \midrule
 \multicolumn{2}{l}{Log-normal particle size distribution:}         & \\
 Median     &13.4968         &$\mu$m \\
 Sigma     &0.2253         &- \\
 Minimum cutoff radius     &10.1117         &$\mu$m \\
 Maximum cutoff radius     &44         &$\mu$m \\
 \bottomrule
 \end{tabular}
\end{table}

\end{document}